\documentclass[prd,nofootinbib,twocolumn,superscriptaddress]{revtex4}
\usepackage{graphics}
\usepackage{color}
\usepackage{bm}

\begin{document}

\title{Zero modes on cosmic strings in an external magnetic field
}

\author{Francesc Ferrer}
\affiliation{
CERCA, Department of Physics, Case Western Reserve University,
10900 Euclid Avenue, Cleveland, OH 44106-7079, USA.}
\author{Harsh Mathur}
\affiliation{
Department of Physics, Case Western Reserve University,
10900 Euclid Avenue, Cleveland, OH 44106-7079, USA.}
\author{Tanmay Vachaspati}
\affiliation{
CERCA, Department of Physics, Case Western Reserve University,
10900 Euclid Avenue, Cleveland, OH 44106-7079, USA.}
\author{Glenn Starkman}
\affiliation{
CERCA, Department of Physics, Case Western Reserve University,
10900 Euclid Avenue, Cleveland, OH 44106-7079, USA.}
\affiliation{BIPAC, Astrophysics, Denys Wilkinson Building, 
Oxford University, Keble Rd, Oxford, UK OX2 3RH.}


\begin{abstract}
\noindent
A classical analysis suggests that an external magnetic field can 
cause trajectories of charge carriers on a superconducting domain 
wall or cosmic string to bend, thus expelling charge carriers with 
energy above the mass threshold into the bulk. We study this process 
by solving the Dirac equation for a fermion of mass $m_f$ and charge 
$e$, in the background of a domain wall and a magnetic field of 
strength $B$. We find that the modes of the charge 
carriers get shifted into the bulk, in agreement with classical 
expectations. However the dispersion relation for the zero modes 
changes dramatically --- instead of the usual linear dispersion 
relation, $\omega_k =k$, the new dispersion relation is well fit by 
$\omega \approx m_f ~{\rm tanh}(k/k_*)$ where $k_*=m_f$ for a thin 
wall in the weak field limit, and $k_*=eBw$ for a thick wall of width $w$.
This result shows that the energy of the charge carriers on the domain 
wall remains below the threshold for expulsion even in the presence of 
an external magnetic field. If charge carriers are expelled due to an 
additional perturbation, they are most likely to be ejected
at the threshold energy $\sim m_f$.
\end{abstract}

\maketitle

\section{Introduction}
\label{introduction}

Fermion zero modes on domain walls and cosmic strings have been
known to exist since the 60's \cite{Caroli:1964,Jackiw:1981ee}.
If the fermions are electrically charged, 
it is generally presumed that the string or wall is superconducting
in the sense that it can sustain long-lived persistent currents
\cite{Witten:1984eb}. Such 
superconducting strings could be present in the universe, and 
would have observable electromagnetic and other signatures 
(for a review, see \cite{VilShe}). Light superconducting strings 
could also be trapped in the center of our galaxy and could explain
\cite{Ferrer:2005xv} the observed positron annihilation line
\cite{SPI}.

The observable signatures of cosmic superconducting strings depend 
on the magnitude of the currents they carry. As the strings move
under their tension, they cut across ambient astrophysical magnetic
fields and this generates currents on the strings. Interactions
of charge carriers on the string are expected to limit the
maximum current that a string can carry~\cite{Barr:1987ij}. 
Hence it is important
to understand the interaction of a superconducting string with
an external magnetic field.

\begin{figure}
\scalebox{0.50}{\includegraphics{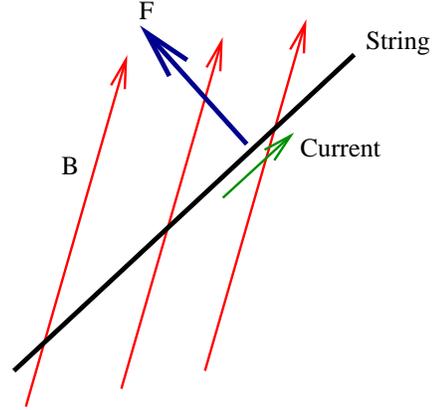}}
\caption{The charge carriers run along the string. In the presence
of a perpendicular external magnetic field, the Lorentz force would
tend to bend the charges off the string.
}
\label{fig.setup}
\end{figure}

In this paper we consider a straight, static, current-carrying, 
string, placed in a background magnetic field. The magnetic field 
is assumed to be perpendicular to the string. The charge carriers 
on the string feel the background magnetic field via the Lorentz 
force which is also perpendicular to the string. Thus the ambient 
magnetic field tends to bend the charge carriers off the string 
as depicted in Fig.~\ref{fig.setup}. 

The dispersion relation for zero modes on a string (with no
external magnetic field) is that for a massless particle
\begin{equation}
\omega_k = +k
\label{omegakzero}
\end{equation}
where, to be specific, we have taken the $+$ sign. (If the
zero mode propagates in the opposite direction along the
string, we would have $\omega_k = -k$.) The dispersion relation 
for the remaining (``non-zero'') modes is of the form
\begin{equation}
\omega_k = \pm \sqrt{{\bf k}^2 + m^2}
\label{omegakmassive}
\end{equation}
where $m$ is a mass associated with the mode. For scattering states,
$m$ is the mass of the fermion in vacuum, $m_f$. For a particle in
a zero mode state to escape, it has to transition into a scattering 
state. Since a particle in a zero mode only has kinetic energy,
to escape into the bulk, its kinetic energy has to transform into 
mass energy. For $k > m_f$, this is not forbidden by energy conservation
because it is possible to find the scattered momentum, ${\bf k}'$,
such that $|k|=\sqrt{{{\bf k}}^{\prime 2} + m_f^2}$. 

Let us consider a particle in a zero mode with 
$k/m \rightarrow \infty$. The particle certainly has enough energy 
to escape the string. The question is whether an external magnetic
field can trigger this escape. 
At first sight it would appear that the fermion zero mode
is very fragile since it is a bound state as regards its
motion perpendicular to the string, yet it is degenerate with
continuum states that are unbound. Such bound states are known
to dissolve resonantly into the the continuum under the effect
of even a very weak perturbation, for example, in the
auto-ionization of excited atomic Helium. However, as shown by
Fano \cite{Fano}, in order to destroy a bound state a perturbation must
couple it to continuum states with which it is degenerate.
Due to the translational symmetry of the problem it is easy to
see that a uniform magnetic field does not couple the zero mode
to continuum states that are degenerate with
it\footnote{The situation is different if the external
magnetic field is non-uniform. However, in an astrophysical
context, the non-uniformity occurs on astrophysical
scales and the ejection rate is correspondingly small.
Other effects that can eject zero modes, such as string curvature,
also occur on these scales. Here we will only be interested
in whether there are effects that can cause faster ejection
than that given by astrophysical scales.}. One way to
understand this result is to note that all the energy of the zero mode is
kinetic energy, while only a fraction of the particle energy
off the string is kinetic. To be ejected from the string,
some of the kinetic energy of the zero mode has to get converted
into mass energy, and the magnitude of the momentum of the
particle off the string therefore has to be less than that of the zero
mode. Thus we are lead to the conclusion that at least perturbatively
a uniform magnetic field does not destroy the zero mode by Fano's
mechanism. It is therefore
necessary to analyse the effect of the magnetic field non-perturbatively
in order to definitively settle the question. In this paper we carry out
such a non-perturbative analysis.

For the reader who may not wish to go through all the details, 
we briefly summarize our result for the dispersion relation 
for fermions on a string or domain wall in the presence of a 
magnetic field.  The dispersion relation for small wavevectors, 
$k$, resembles that of a zero mode on a string without any
magnetic field. Intuitively, $k$ is the momentum 
of the particle and at zero momentum, there is no Lorentz force 
and so the magnetic field plays no role. Hence the dispersion relation
is linear for small $k$. On the other hand, at large $k$, the 
magnetic field dominates and the dispersion relation is as if 
there is no domain wall. In this regime, the energy is independent 
of $k$, just as in the Landau problem. The transition of the dispersion 
from linear in $k$ to independent of $k$ occurs when the effect of 
the domain wall is comparable to that of the external magnetic field, 
or approximately at $k_*$ where
$k_*=m_f$ for a thin 
wall in the weak field limit, and $k_*=eBw$ for a thick wall of width $w$.
Thus the fermions in the lowest energy level have a dispersion 
relation that is of the form
\begin{equation}
\omega_k = m_f {\rm tanh} \biggl ( \frac{k}{k_*} \biggr )
\label{analyticdisp}
\end{equation}
This dispersion is exact in the thick wall limit and fits our numerical 
results reasonably well in other cases.

The modified form of the dispersion relation for superconducting 
strings may be important in astrophysical applications as briefly
described in the concluding section.

\section{Simplifying the problem}
\label{simplify}

The proper framework for discussing the effect of external
magnetic fields on zero modes on strings is in terms of the
field theory of scalar and gauge fields that gives a string
solution, the fermions which have zero modes on the string, 
and the external magnetic field. Such a model is quite
complicated and it is desirable to find a simpler model
with all the relevant physics. For this reason we consider
a field theory in two spatial dimensions that has domain
wall solutions, a fermion with zero modes on the domain
wall, and a magnetic field. The Lagrangian for the model
containing a scalar field, $\phi$, a fermion, $\psi$, and
a vector field, $A_\mu$, is
\begin{eqnarray}
L &=& \frac{1}{2} (\partial_\mu \phi)^2 + 
     \frac{m^2}{2} \phi^2 - \frac{\lambda}{4}\phi^4
     -\frac{1}{4} F_{\mu\nu} F^{\mu\nu} \nonumber \\
     &+&i {\bar \psi} \gamma^\mu (\partial_\mu + ie A_\mu )\psi
     - g \phi {\bar \psi}\psi
\end{eqnarray}
The $\gamma$ matrices are defined, in terms of the Pauli spin matrices, by
\begin{equation}
\gamma^t = \sigma_3 
\ , \ \ 
\gamma^x = i \sigma_2 
\ , \ \ 
\gamma^y = -i \sigma_1 
\end{equation}

The scalar sector of the field theory is the $\lambda \phi^4$
model and has the classical domain wall solution
\begin{equation}
\phi (x) = \eta {\rm tanh} \left ( \frac{x}{w} \right )
\label{wallsolution}
\end{equation}
where $\eta = m/\sqrt{\lambda}$ and $w = \sqrt{2}/m$ is 
the width of the wall.

In the absence of an external magnetic field, the fermionic
zero mode is
\begin{equation}
\psi_0 (x)= N \left[{\rm sech}\left(\frac{x}{w}\right)\right]^{g \eta w}
\left(\begin{array}{c} 1 \\ i 
\end{array}\right) 
\end{equation}
This solution has zero momentum and zero energy. It can be
boosted along the direction of the wall to get a propagating
zero mode
\begin{equation}
\psi_k (t,x,y) = \exp \left[{-i (\omega t -k y)}\right] \psi_0 (x)
\label{factorpsi}
\end{equation}
The dispersion relation is as in Eq.~(\ref{omegakzero})
\begin{equation}
\omega^0_k = + k
\end{equation}

In Sec.~\ref{wallinB} we will discuss the zero modes when we also 
have an external magnetic field. To gain intuition for the final
results, we now discuss the Landau problem of fermions in a magnetic 
field but without a domain wall.

\section{Relativistic Landau problem}
\label{landau}

We now consider the fermion modes in a magnetic field taken to 
be uniform in the $xy-$plane. The potential is
\begin{equation}
A^\mu = \left ( 0, 0, - B x \right ) \ , \ \
\mu = t,x,y
\label{gaugefield}
\end{equation}

With this choice of gauge, the fermion wavefunctions can be
factorized as in Eq.~(\ref{factorpsi}) and we can solve for
the functions $\psi^B$. 

There is a special eigenmode with $\omega_{0,k}^B = -m_f$
\begin{equation}
\psi^B_{0,k} = \left(\begin{array}{c} 0 \\ \varphi_0 \left(x+x_0\right)
\end{array}\right) e^{-i(\omega^B_{0,k} t -ky)}
\label{zerolandau}
\end{equation}
where 
\begin{equation}
x_0 = \frac{k}{eB}
\end{equation}
The remaining relativistic Landau eigenfunctions are
\begin{eqnarray}
\psi^B_{n,k\pm} &=& A_{n,\pm} 
\left(\begin{array}{c} \varphi_n \left(x+x_0\right)\\ \\
\frac{i \sqrt{2 e B (n+1)}}{m_f \pm \sqrt{m_f^2 + 2 e B (n+1)}} 
\varphi_{n+1} \left(x+x_0\right)
\end{array}\right) \nonumber \\
 && \hskip 1in \times e^{-i(\omega^B_{n,k,\pm} t-ky)}
\end{eqnarray}
with energy eigenvalues
\begin{equation}
\omega^B_{n,k,\pm} = \pm \sqrt{m_f^2 + 2 e B (n+1)}, \quad n=0,1,2,\ldots
\label{dispersionlandau}
\end{equation}
Note that the energy eigenvalues in the Landau
problem are independent of the wavevector $k$.

Here $A_{n,\pm}$ are normalization constants and $\varphi_n(x)$ 
is the solution to the shifted harmonic oscillator equation which 
can be expressed in terms of Hermite polynomials:
\begin{equation}
\varphi_n(x)=\frac{(e B)^{1/4}}{\pi^{1/4} 2^{n/2}n!} H_n \left( x \sqrt{e B} 
\right) \exp \left(\frac{-e B x^2}{2} \right)
\label{hermitelandau}
\end{equation}

\section{Domain wall in a magnetic field}
\label{wallinB}

\begin{figure}
\centerline{\scalebox{1.00}{\input{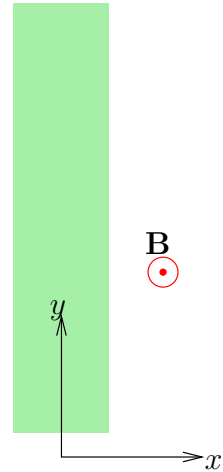}}}
\caption{In our simplified setup, a domain wall lies along 
the $y-$axis and a uniform magnetic field lies in the 
$xy-$plane. Fermion zero modes run along the wall {\it i.e.}
in the $y-$direction.
}
\label{fig.wallsetup}
\end{figure}

Now we consider the full setting where there is a domain
wall along the $y-$axis, and a uniform magnetic field of magnitude 
$B$ in the $xy-$plane (see Fig.~\ref{fig.wallsetup}). Then the 
scalar field is as given in Eq.~(\ref{wallsolution}) and the 
gauge field in Eq.~(\ref{gaugefield}).

The Dirac equation now reads
\begin{equation}
\left( i \gamma^\mu \partial_\mu + e B x \gamma^2 - g \phi(x) \right)\psi = 0
\label{fullDiraceq}
\end{equation}
This is the equation for a Dirac particle moving in a uniform magnetic field
with a position dependent mass $m(x)=g \phi(x)$. Far from the wall, the mass
approaches the asymptotic value $\pm m_f$ as $x \rightarrow \pm \infty$, 
where $m_f = g \eta$ 

To find the eigenstates of Eq.~(\ref{fullDiraceq}), we write
\begin{equation}
\psi (t,x,y) = \pmatrix{\chi_1 (x)\cr \chi_2 (x)} e^{-i(\omega t -k y)}
\end{equation}
The equation for the $\chi$ fields is 
\begin{equation}
\left( \begin{array}{cc} 
m(x)-\omega & -i \partial_x -i k -i e B x \\
-i \partial_x +i k +i e B x & -m(x)-\omega
\end{array} \right) 
\left( \begin{array}{c} \chi_1 \\ \chi_2 \end{array} \right) 
         = 0
\label{chieq}
\end{equation}
The dimensionful scales that characterize the system are: $w$, 
$m_f$, and $e B$, with $w$ and $m_f$ both occurring 
within $m(x)$. These dimensionful parameters can be written as 
three length scales in the problem: $w$, 
$l_\psi \equiv m_f^{-1}$ and $l_B \equiv 1/\sqrt{eB}$. 
In addition, a mode is characterized by its wavelength
$2\pi /k$. Only the ratio of the various length scales 
is important.  For convenience, we work in units where $w=1$
except in the thin wall approximation (Sec.~\ref{thinwall})
where we take $w \rightarrow 0$.

We would like to find all solutions to Eq.~(\ref{chieq}).
This is possible to do analytically in two extreme limits
(1) thick domain wall ($w \gg l_\psi ,\: l_B$), 
and (2) thin domain wall ($w \ll l_\psi ,\: l_B$). 
In general, the modes need to be found numerically.
Before discussing the solutions, it is necessary to clarify
the concept of a zero mode in the presence of a magnetic field
and to prove that such a mode does exist in general and not 
only in the two limiting cases for which we can find exact 
solutions.

\subsection{Zero mode at $k=0$}


At zero magnetic field, the key distinguishing features of the 
zero mode state are that it is localised in a direction perpendicular
to the domain wall (along the $x$-axis) and has zero energy at 
$k=0$. With a magnetic field turned on, in the present
gauge, all wavefunctions are localised along the
$x$-direction, but we can still ask whether there is a zero
energy state and, if so, identify it as the magnetic field
analog of the zero-mode state. To this end, we must study
Eq.~(\ref{chieq}) and ask whether it has well-behaved
solutions with $\omega = 0$ when we set $k=0$.

In the zero field case studied in section \ref{simplify},
we were able to establish the existence of the zero mode by
explicit, exact solution of the Dirac equation. This is a rare
luxury; under more typical circumstances it is necessary to
resort to index theorems \cite{eweinberg} or other indirect
arguments to establish the existence of zero modes. The
strategy we follow here is to eliminate $\chi_2$ and rewrite
Eq.~(\ref{chieq}) as a second-order equation
\begin{eqnarray}
&&- \frac{ \partial^2 }{\partial x^2} \chi_1 +
\frac{m'}{m} \frac{\partial}{\partial x} \chi_1 + \nonumber \\
&&\left( m^2 + e B + e^2 B^2 x^2 -
\frac{m'}{m} e B x \right) \chi_1 = 0,
\label{eq:secondorder}
\end{eqnarray}
where $m \equiv m(x)$ and $m' \equiv \partial m/ \partial x$.
Once this equation has been solved for $\chi_1$ we may
determine $\chi_2$ via
\begin{equation}
\chi_2 = - \frac{i}{m} \frac{\partial}{\partial x} \chi_1 +
i \frac{eB}{m} x \chi_1
\label{eq:chitwo}
\end{equation}
For $ x \rightarrow \infty$ we may set $m' = 0$ in Eq.
(\ref{eq:secondorder}) and thereby find that in this asymptotic region
$\chi_1$ obeys the Schr\"{o}dinger equation of a non-relativistic
oscillator. Therefore it has one solution that grows as $x \rightarrow \infty$
and one that is well-behaved and decays. The key question now is whether
a well-behaved solution can be found at $x=0$ that will connect with the
well-behaved solution at $+ \infty$. Since $m(x) = 0$ for $x=0$,
it is clear that the origin is a (regular) 
singular point for the differential equation 
(\ref{eq:secondorder}) and it appears to be a possibility
that of the two independent solutions near the origin only one
is regular. However, by use of a series expansion, we can show
that there are in fact two independent solutions that are both regular
at $x=0$. One has the series expansion
\begin{equation}
\chi_1 = 1 + \frac{1}{8} \left(
\mu_1^2 - 2 \frac{\mu_3}{\mu_1} e B + e^2 B^2 \right) x^4 + \ldots;
\label{eq:frobeniusone}
\end{equation}
the other has the expansion
\begin{equation}
\chi_1 = x^2 + \frac{\mu_3}{2 \mu_1} x^4 + \ldots
\label{eq:frobeniustwo}
\end{equation}
Here we have taken $m(x) = \mu_1 x + \mu_3 x^3 + \ldots$ near the origin.
A generic combination of these two solutions would develop into
a superposition of the growing and decaying solutions as
$x \rightarrow \infty$; but by fine-tuning the coefficients we can
arrange for the solution to decay as $x \rightarrow \infty$.
Furthermore, since the solutions in Eqs.~(\ref{eq:frobeniusone})
and (\ref{eq:frobeniustwo}) are both even, and the differential
Eq.~(\ref{eq:secondorder}) is also even, it follows that the solution
will be even and well-behaved as $ x \rightarrow - \infty$ as well.
This concludes our demonstration that there is a zero energy solution
to Eq.~(\ref{chieq}) for $k=0$.

\subsection{Zero mode for non-zero $k$}

We turn now to the magnetic zero mode for non-zero $k$.
It is shown below that for a fixed $k$ Eq.~(\ref{chieq}) 
has a discrete spectrum of eigenvalues $\omega$. Thus we may hope to
identify the zero mode at non-zero $k$ as the state that smoothly
evolves into the zero energy state as $k
\rightarrow 0$. That such a state exists
is not obvious: there is a competing possibility that the spectrum is
not continuous as $k \rightarrow 0$ and that the zero energy state at
$k=0$ has no counterpart for non-zero $k$. In this section, however,
we will give a variational argument that makes it certain beyond
reasonable
doubt that the spectrum {\em is} continuous at $k=0$ and that the zero
mode
therefore continues to exist for non-zero $k$. Further, we are able to
verify
this continuity and to study the evolution of the zero mode with
$k$ explicitly in the thick and thin wall limits discussed in the
subsections
below.

To study the magnetic zero mode for non-zero $k$ it is helpful to rewrite
Eq.~(\ref{chieq}) as a second-order equation by analogy to
Eq.~(\ref{eq:secondorder}). We obtain
\begin{eqnarray}
\frac{ \partial^2 }{ \partial x^2 } \chi_1 &-&
\frac{m'}{(\omega + m)}
\frac{ \partial }{ \partial x } \chi_1 +
\left[ \omega^2 - m^2 - e B - (k + e B x)^2 \right. \nonumber \\ &+&
\left.
\frac{m'}{(\omega + m)} (k + e B x) \right] \chi_1 = 0
\label{eq:secondordergeneral}
\end{eqnarray}
where we have used
\begin{equation}
\chi_2 = - \frac{i}{(\omega + m)} \frac{\partial}{\partial x} \chi_1
+ \frac{i e B x}{(\omega + m)} \chi_1 + i k \chi_1
\label{eq:chitwogeneral}
\end{equation}
to eliminate $\chi_2$ from the coupled Eq.~(\ref{chieq}).
As $ x \rightarrow \pm \infty$ we may set $ m' = 0 $ and find that
Eq. (\ref{eq:secondordergeneral}) turns into the equation of a
non-relativistic oscillator. Therefore it has a discrete spectrum.

For $\omega$ outside the mass gap ($\omega > m_f$ or $\omega < - m_f$)
Eq. (\ref{eq:secondordergeneral}) is not singular on the real axis.
Thus for any $\omega$ we can find a solution that is well behaved
at the origin and as $x \rightarrow \infty$. By tuning $\omega$ to
a quantised energy level we can also ensure that it is well behaved
as $x \rightarrow - \infty$. Thus it is easy to see that a discrete
spectrum will arise outside the mass gap. However for the
present purpose these states are of no interest since they will not
continuously evolve into the zero energy state as $k \rightarrow 0$.

If we look for states that lie below the mass gap ($- m_f < \omega < m_f$)
we must come to grips with the fact that Eq.~(\ref{eq:secondordergeneral})
is singular at $x_0$ where $x_0$ is fixed by the condition $ \omega +
m(x_0)
= 0$. By making a series expansion about this point it is easy to show
that Eq.~(\ref{eq:secondordergeneral}) has one well behaved solution at
this point and one that is singular. Thus we must fine-tune $\omega$ in
order to obtain a solution that behaves well as $x \rightarrow
\infty$\footnote{More precisely, we find that both solutions for
$\chi_1$ are regular at $x_0$, but when we compute $\chi_2$ using
Eq. (\ref{eq:chitwogeneral}) we find that one of them leads to a
singular $\chi_2$. By tuning $\omega$ we can ensure that both solutions
at $x_0$ are regular and hence that the solution at $ x \rightarrow
\infty$
is well behaved. Alternatively we can accept that only one solution is
regular at $x_0$. For generic $\omega$ this solution will correspond
to a superposition of growing and decaying solutions as $x \rightarrow
\infty$; but by tuning $\omega$ we may expect to be able to arrange for
the
solution regular at $x_0$ to evolve into one that is well behaved at
$x \rightarrow \infty$.}. Since Eq.~(\ref{eq:secondordergeneral}) is not
symmetric about $x_0$, it is not obvious that the solution will be
well behaved as $x \rightarrow - \infty$. This raises the real concern
that there may be no eigenstates within the mass gap for non-zero $k$
and therefore no continuation of the $k=0$ state of zero energy.

To put this concern to rest and to establish that there are in fact
states within the mass gap we turn to a variational argument.
To this end, it is convenient to write Eq.~(\ref{chieq}) in the
form
\begin{equation}
H \left(
\begin{array}{c}
\chi_1 \\
\chi_2
\end{array}
\right) = \omega \left(
\begin{array}{c}
\chi_1 \\
\chi_2
\end{array}
\right).
\label{eq:hform}
\end{equation}
Here $H$ is an hermitean matrix differential operator and we regard
$k$ as a fixed parameter. Now $H^2$ has a spectrum that is bounded
from below and it is therefore an operator to which the variational
principle may be applied. We will seek a normalised state 
$|\Omega\rangle$ that satisfies 
$ \langle \Omega| H^2 | \Omega\rangle < m_f^2 $. This implies that 
$H^2$ has an eigenvalue below $m_f^2$ for every value of $k$, 
and in turn that $H$ has an eigenvalue within the mass 
gap\footnote{Since $[H,H^2]=0$, eigenstates of $H^2$ can be
chosen to be eigenstates of $H$ and the eigenvalues 
of $H^2$ are simply the squared eigenvalues of $H$.}. 

For the state $| \Omega \rangle $ we take
\begin{equation}
\left(
\begin{array}{c}
\chi_1 \\
\chi_2
\end{array}
\right) =
\frac{1}{\sqrt{2}} \left(
\begin{array}{c}
\exp ( - |x| ) \\
i \exp ( - |x| )
\end{array}
\right).
\label{eq:trial}
\end{equation}
A straightforward computation reveals
\begin{eqnarray}
\frac{1}{m_f^2} \langle \Omega | H^2 | \Omega \rangle &=&
3 - 4 \ln 2 - \frac{4}{m_f} \left( \ln 2 - \frac{1}{2} \right) \nonumber \\
&&+ \frac{1}{m_f^2} \left( 1 + k^2 + \frac{1}{2} e^2 B^2 \right).
\label{eq:trialenergy}
\end{eqnarray}
We would like to ensure that the right hand side of Eq.~(\ref{eq:trialenergy})
is less than unity. For suitable parameters this is clearly the case.
Thus at least over a certain range of $m_f$ and $eB$ there are sub-gap
states for small non-zero $k$. By continuity we may therefore reasonably
conclude that quite generally there is a state that is the continuation
to non-zero $k$ of the zero energy state at $k=0$. We will confirm this
expectation by approximate solution to the problem in the thick wall limit
and exact solution in the thin wall limit below.

\subsection{Thick wall}

In the limit that the wall thickness is large it is
possible to derive asymptotic expressions for the zero mode
as well as the lowest few excited states that lie outside the
mass gap. To understand the approximation involved it is instructive
to set $m' \rightarrow 0$ in Eq.~(\ref{eq:secondordergeneral}).
On doing so we see that $\chi_1$ obeys the Schr\"{o}dinger equation
for a harmonic oscillator centered at $x_0 = - k/eB$. The width of the
oscillator wave function is $1/\sqrt{eB}$. Since we are working in units
where the wall width $w=1$, it would appear that the approximation
of treating $m(x)$ as constant and equal to $m(x_0)$ should be
accurate when the extent of the wavefunctions is small compared to
the distance over which the mass varies, {\em i.e.}, in the
high field regime $1/(eB) \ll 1$. In Appendix~\ref{appa} we
will show that the condition for validity is slightly more restrictive.

In summary the thick wall approximation consists of replacing
$m(x)$ in Eq.~(\ref{chieq}) with the constant $m(x_0)$ where
$ x_0 = - k/(eB)$. The solution to Eq.~(\ref{chieq})
in this approximation may then be read off from the solution to
the relativistic Landau problem given in section~\ref{landau}.
Proceeding in this way, we find that the zero mode state that
lies within the mass gap has an energy
\begin{equation}
\omega_B^{0} (k) = - m(x_0) = m_f \tanh \left( \frac{k}{eB} \right)
\label{eq:thickdispersion}
\end{equation}
and a wavefunction given by Eqs.~(\ref{zerolandau}) and
(\ref{hermitelandau}).

There are several noteworthy features of this zero-mode solution.
First observe that indeed $ \omega_B^{0} \rightarrow 0$ as $k
\rightarrow 0$, showing that this state is continuous with
the zero energy state at $k=0$. Hence this state is identifiable
as the zero mode state for non-zero $k$. Next we see that in
contrast to the situation at zero magnetic field, the dispersion
relation of the zero mode is linear only for small $k$. For
larger $k$, $\omega_B^{0}$ saturates and always remains within
the mass gap even as $k \rightarrow \pm \infty$. At zero field
the zero mode wave function is always localised about the domain wall;
indeed it is independent of $k$ in this case. By contrast with a magnetic
field the zero mode wavefunction is centered about $x_0 = - k/(eB)$
and it moves far away from the domain wall as $k \rightarrow \pm \infty$.
The shape of the wavefunction is always a Gaussian and does not
depend on $k$.

Evidently in the thick wall limit the low lying excited states
of Eq.~(\ref{chieq}) are also given by Eqs.~(\ref{dispersionlandau}),
 and (\ref{hermitelandau}) if we substitute
$ m \rightarrow m(x_0) = - m_f \tanh (k/eB)$. For sufficiently
excited states this approximation will break down because the
width of the wavefunctions grows to a point that it is no
longer appropriate to treat $m(x)$ as a constant equal to its
value at the center of the wavefunction.

To conclude we give the conditions for validity of the thick
wall approximation to the zero mode. It is shown in Appendix~\ref{appa}
that the approximate solution is good if $eB \gg 1$ and either
$|x_0| = |k/eB| \gg 1$ or $m_f \ll e B$. Here, as elsewhere in
this subsection, we have worked in units where the wall width
$w = 1$.

\subsection{Thin wall}
\label{thinwall}

In this approximation the wall profile corresponds to a sharp step
\begin{equation}
m(x) = m_f \Theta ( x ) - m_f  \Theta ( - x )
\end{equation}
With this profile Eq.~(\ref{eq:secondordergeneral}) leads to the 
following second order equation for $\chi_1$
\begin{equation}
\frac{\partial^2 \chi_1}{\partial x^2}+
\left[\omega^2 - m_f^2 -e B - (k+eBx)^2 \right] \chi_1=0 \ , \ \ 
x \ne 0
\label{thin1}
\end{equation}

This is the equation of a shifted harmonic oscillator.
With the change of variables $z= \sqrt{2 e B} \left(x+x_0 \right)$, where
$x_0=k/eB$, Eq.~(\ref{thin1}) takes the form of Weber's 
equation~\cite{whittaker}:
\begin{equation}
\frac{\partial^2 \chi_1}{\partial z^2}+
\left[\nu + \frac{1}{2}-\frac{z^2}{4}\right] \chi_1=0
\label{eqweber}
\end{equation}
Here 
\begin{equation}
\nu =\frac{\omega^2-m_f^2}{2 e B}-1
\label{nweber}
\end{equation}
is not generally an integer.

A well-behaved, {\it i.e.} tending to 0 as $x \rightarrow \pm \infty$, 
solution to Eq.~(\ref{eqweber}) can be given in terms of the Weber
function
\begin{equation}
\chi_1(x)=\left\{ \begin{array}{l c} 
D_\nu\left(\sqrt{2 e B} \left(x+x_0 \right)\right), & x > 0 \\
C D_\nu\left(- \sqrt{2 e B} \left(x+x_0 \right)\right), & x < 0 
\end{array} \right.
\label{chi1weber}
\end{equation}
where $\nu$ is given in Eq.~(\ref{nweber}) and 
\begin{equation}
C=\frac{D_\nu \left( x_0\sqrt{2 e B} \right)}
       {D_\nu \left(-x_0\sqrt{2 e B} \right)}
\label{cchi}
\end{equation}
is found by demanding continuity of the wavefunction at the origin.
Note that the derivative $\partial_x \chi_1$ will be discontinuous at
$x=0$ due to the non smooth wall profile there. 
The quantization of $\nu$ in Eq.~(\ref{nweber}) arises in 
the simple harmonic oscillator problem in quantum
mechanics from imposing that $\partial_x \chi_1$ be continuous, but
in the problem at hand, the energy will only be determined after 
requiring continuity of the lower component, $\chi_2$, of the eigenmodes
\begin{equation}
\chi_2(x)=\left\{ \begin{array}{l c} 
\frac{-i}{\omega+m_f} \left[\partial_x \chi_1- (k+e B x) \chi_1 \right]
, & x > 0 \\ & \\
\frac{-i}{\omega-m_f} \left[\partial_x \chi_1- (k+e B x) \chi_1 \right]
, & x < 0 
\end{array} \right.
\label{chi2thin}
\end{equation}
This can also be expressed in terms of Weber functions
\begin{equation}
\chi_2(x)=\left\{ \begin{array}{l c} 
\frac{-i}{\omega+m_f} 
\left[\sqrt{2 e B} \nu D_{\nu-1}-2 (k+e B x) D_\nu \right]
, & x > 0 \\ & \\
\frac{-i C}{\omega-m_f} 
\left[-\sqrt{2 e B} \nu D_{\nu-1}-2 (k+e B x) D_\nu \right]
, & x < 0 
\end{array} \right.
\label{chi2weber}
\end{equation}
where $C$ is also given by Eq.~(\ref{cchi}).

Requiring that $\chi_2$ be continuous leads to the quantization relation 
that determines the possible energies of the eigenmodes:
\begin{eqnarray}
&&4 k m_f D_\nu \left(-k \sqrt{2/eB}\right) D_\nu \left(k \sqrt{2/eB} \right) =
-\sqrt{2 eB} \nu \nonumber \\
&&\times \left[ (\omega+m) D_{\nu-1}\left(-k \sqrt{2/eB}\right) 
D_\nu \left(k \sqrt{2/eB} \right)\right. \nonumber \\
&&+\left.(\omega-m)D_{\nu-1}\left(k \sqrt{2/eB}\right) 
D_\nu \left(-k \sqrt{2/eB} \right)\right]
\label{energyquant}
\end{eqnarray}
where $\nu$ is given in Eq.~(\ref{nweber}). 

In order to numerically solve the condition in Eq.~(\ref{energyquant}),
it is convenient to express the Weber functions in terms of 
hypergeometric functions. This is done in Appendix~\ref{weberhypergeom},
In Fig.~\ref{figdisp} we show the shape of the dispersion relation  
for the zero modes for $e B = m_f=1$ and a few of the scattering
states.

The numerical results from Eq.~(\ref{energyquant}) can be approximately fit
by Eq.~(\ref{analyticdisp}). Indeed, using the results 
in Appendix~\ref{weberhypergeom}, we can study the dispersion relation 
around the origin, $k=0$, to find $k_*$:
\begin{eqnarray}
&&\frac{m_f}{k_*}=\left. \frac{d \omega}{dk} \right|_{k=0} = 
2^{1-v^2} \sqrt{\frac{\pi}{2}} \: v \: \Gamma \left(2+v^2 \right) \times
\nonumber \\
&&\left[\frac{1}{\left(1+v^2 \right) \Gamma \left(1+\frac{v^2}{2} \right)}- 
\frac{1}{\Gamma \left(1+\frac{v^2}{2}\right)^2}+ \right. \nonumber \\
&&\left.
\frac{1}{\Gamma \left(\frac{1}{2} +\frac{v^2}{2} \right) 
\Gamma \left(\frac{3}{2} +\frac{v^2}{2}  \right)} \right],
\label{kstarthin}
\end{eqnarray}
where $v \equiv m_f/\sqrt{2 e B}$. The absolute value of the slope 
$d \omega/ dk$ at the origin is always less than $1$, approaching this 
value asymptotically for large values of $|v|$, {\it i.e.} for small 
magnetic fields.

\begin{figure}
\scalebox{.8}{\includegraphics{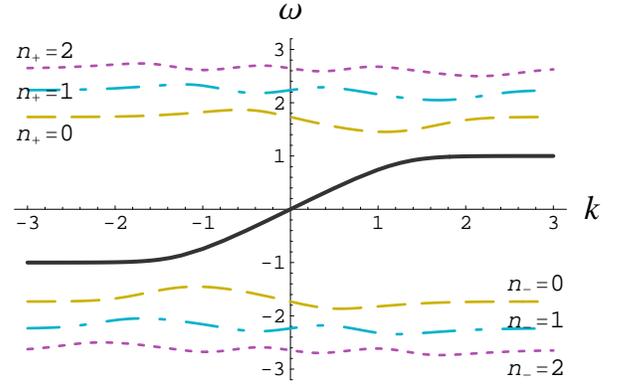}}
\caption{Dispersion relation for the zero modes and the lowest
few excited states that lie outside the mass gap for $e B=m_f=1$.
The integers $n_\pm$ correspond to the integer $n$ in 
Eq.~(\ref{dispersionlandau}) and the $\pm$ sign that occur in
the relativistic Landau problem. Note the symmetry 
$\omega_{n_+}(k)=-\omega_{n_-}(-k)$, which is exact
and not limited to the thin wall approximation.
}
\label{figdisp}
\end{figure}

\section{Conclusions}

We have studied fermionic zero modes on a domain wall with 
an external magnetic field in the configuration shown in 
Fig.~\ref{fig.wallsetup}. This is more tractable than
the case of zero modes on a string in an external magnetic
field (Fig.~\ref{fig.setup}) and is expected to have
similar physics since the essential ingredients of the two
systems are identical.

We have found that the zero mode continues to survive
even in an external magnetic field but is now centered
off the wall. 
This is perhaps not so surprising because even at the classical 
level the magnetic force tends to deflect the fermion away from
the wall.
On the other hand, we are accustomed to thinking of the
zero mode as being localized on the wall and, in the
external magnetic field, this is certainly not the
case.

\begin{figure}
\scalebox{.45}{\includegraphics{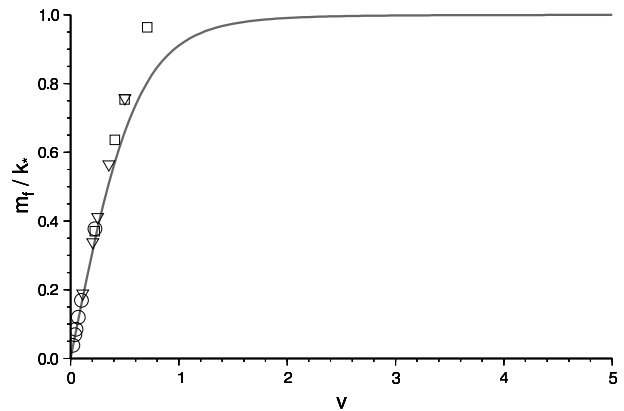}}
\caption{
The slope, $\left. d \omega /dk \right|_{k=0} \equiv m_f/k_*$, as 
a function of $v \equiv m_f/\sqrt{2 e B}$. The symbols show $m_f/k_*$ 
as found by a global fit to Eq.~(\ref{dispfit}) for 
$m_f = 0.1, \; 0.5 \;{\rm and}\;1$ (circles, triangles and squares). 
The curve is a plot of Eq.~(\ref{kstarthin}) and there is good 
agreement with the global fit for small $v$ and for all $m_f$. 
The deviation at large $v$ may be due to numerical issues and 
departures from the fitting formula in Eq.~(\ref{dispfit}). 
}
\label{dispersionfit}
\end{figure}

In the absence of an external magnetic field, the dispersion
relation for the zero modes is linear and the energy
increases indefinitely with increasing momentum. In the
presence of the external magnetic field, this picture 
changes dramatically and the dispersion relation has the
shape shown in Fig.~\ref{figdisp}. The form
of the dispersion relation for the zero mode can be approximated 
by 
\begin{equation}
\omega_k = m_f {\rm tanh}\left ( \frac{k}{k_*} \right )
\nonumber
\label{dispfit}
\end{equation}
where, in the
case of a thick wall, Eq.~(\ref{eq:thickdispersion}) shows that 
this approximation is exact with $k_* = e Bw$, 
$w$ being the wall thickness. For the thin wall, the fit is more accurate for
small values of $v$, as shown in Fig.~\ref{dispersionfit}, where it is also
shown that $k_*$ can be approximated by its exact value at the origin 
given by Eq.~(\ref{kstarthin}). 
Note that $| \omega_k | < m$ for all $k$. 
It is also worth reiterating that the dispersion relation is linear
only for $k < k_{*}$. For larger $k$, the dispersion saturates
and hence charge carriers actually propagate with a lower group 
velocity, $d \omega_k/ d k$, which vanishes exponentially fast
with increasing $k$.
In the astrophysical context magnetic fields are weak in the
sense $m_f \gg \sqrt{2 e B}$
and so the crossover scale $k_{*} \approx m_f$. 

For a fermion in a zero mode to escape the wall, it necessarily 
requires some additional energy to transition from the zero mode 
to an energy larger than $m_f$. The additional energy can be 
exponentially small for $k > k_* $ and hence these fermion 
zero modes are extremely fragile -- even a small perturbation
can knock out the particles in these modes. For weak perturbations,
the particles will be ejected into the lowest non-zero ($n\ne 0$)
modes which also have energy $\sim m_f$. The perturbation itself
can be inhomogeneities in the magnetic field, string motion
and curvature, or the interactions of particles outside and
on the string.

The modified dispersion has consequences for the cosmology
of superconducting strings, including the formation of
stable vortons \cite{VilShe}. In connection with the scenario 
of positron production by superconducting strings in the Milky 
Way \cite{Ferrer:2005xv}, our analysis indicates that positrons 
will be produced at an energy comparable to their rest mass, or 
about 511 keV. Then the annihilation of these positrons with ambient 
electrons will also produce 511 keV $\gamma-$rays.

\begin{acknowledgments} 
This work was supported by the U.S. Department of Energy, 
the National Science Foundation, and NASA at Case Western Reserve 
University. GDS was supported in part by fellowships from the John 
Simon Guggenheim Memorial Foundation, the Beecroft Institute for 
Particle Astrophysics and Cosmology, and The Queen's College, Oxford.
\end{acknowledgments}

\appendix
\section{Validity of the Thick Wall Approximation}
\label{appa}

We now evaluate the leading corrections to the thick wall
approximation used in section IVC. To this end it is helpful
to write the Hamiltonian in Eq.~(\ref{chieq}) in the form
$H = H_0 + V$. Here $H_0$ is the Hamiltonian with the replacement
$m(x) \rightarrow m(x_0)$ where $x_0 = -k/(eB)$. The perturbation
\begin{equation}
V = \left(
\begin{array}{cc}
m(x) - m(x_0) & 0 \\
0 & m(x_0) - m(x)
\end{array}
\right)
\label{eq:pert}
\end{equation}
may be further decomposed as
$V = V_1 + V_2 + \ldots $ where
\begin{eqnarray}
V_1 & = &
m'(x_0) (x - x_0) \left(
\begin{array}{cc}
1 & 0 \\
0 & -1
\end{array}
\right)
\nonumber \\
V_2 & = &
\frac{1}{2} m''(x_0) (x - x_0)^2
\left(
\begin{array}{cc}
1 & 0 \\
0 & - 1
\end{array}
\right).
\label{eq:vonevtwo}
\end{eqnarray}
In the thick wall limit we expect that $m(x)$ is
slowly varying so that $V_2$ is nominally smaller
than $V_1$. However we will find that both perturbations
have a comparable effect on the energy of the zero mode
state.

The zeroth-order solution to $H$ is the thick wall
approximation presented in section IVC. The perturbation
$V_1$ does not perturb the energy of the zero mode to first
order since $\langle {\rm zero} | V_1 | {\rm zero} \rangle = 0$.
A straightforward calculation shows that the second order correction
due to $V_1$ is
\begin{equation}
\omega^{(2)}_{{\rm zero}} = - \frac{ m_f^3}{2 e^2 B^2} \tanh x_0
{\rm sech}^4 x_0.
\label{eq:secondorderpert}
\end{equation}
To assess the accuracy of the zeroth-order solution, we note that
\begin{equation}
m_f - \omega_B^0 \rightarrow 2 m_f \exp( - 2 x_0 ) \ , \ \ 
                  x_0 \rightarrow \infty
\end{equation}
On the other hand the correction 
\begin{equation}
\omega^{(2)}_{{\rm zero}} 
\rightarrow - m_f^3 \exp( - 4 x_0 )/(2 e^2 B^2) \ , \ \ 
                  x_0 \rightarrow \infty 
\end{equation}
Thus the perturbation is far smaller than the distance to the mass 
gap and may be considered negligible for $x_0 \gg 1$. We would also 
like to ensure that the perturbation $\omega^{(2)}_{{\rm zero}}$ 
does not shift the zero mode energy outside the mass gap for 
$x_0 \sim 1$ or $x_0 \ll 1$. Therefore our criterion becomes
$| \omega^{(2)}_{{\rm zero}} | \ll m_f$, which leads to the condition
$m_f \ll e B$. If $V_1$ was the only perturbation we would conclude that
the zeroth-order approximation is accurate when $x_0 \gg 1$ or when
$ m_0 \ll eB$.

However, we must also take into account the perturbation $V_2$.
Although nominally smaller than $V_1$, $V_2$ perturbs the zero mode
energy to first order and therefore has a more significant effect.
The first order correction is
\begin{equation}
\omega^{(1)}_{{\rm zero}} = - \frac{m_f}{2 e B} \exp( - 2 x_0 ).
\label{eq:firstorder}
\end{equation}
Applying the same reasoning as for the second order correction
we conclude that the perturbation $V_2$ is insignificant when
either (i) $ x_0 \gg 1$ and $ eB \gg 1$ or (ii) $ e B \gg m_f$
and $ e B \gg 1$. 
These conditions subsume the conditions for
$V_1$ to be negligible and are therefore the conditions that must
be satisfied for the validity of the thick wall approximation to
the zero mode.

\section{From Weber to hypergeometric function}
\label{weberhypergeom}

We can find the dispersion relation of the zero modes for a given 
magnetic field by finding the roots of Eq.~(\ref{energyquant}) 
that satisfy $\omega < m_f$ for different values of $k$.
For this purpose it is useful to consider the expansion of the Weber
function in the complex plane in terms of the Kummer confluent 
hypergeometric function ${}_1F_1(a;\rho;z)$: 
\begin{eqnarray}
&&D_\nu(z)=\sqrt{\pi}2^{\nu/2} e^{-z^2/4} \times \nonumber \\ && \left\{ 
\frac{ {}_1F_1 \left( -\frac{\nu}{2};\frac{1}{2};\frac{z^2}{2} \right)}
{\Gamma \left(\frac{1}{2}-\frac{\nu}{2} \right)}-
\sqrt{2} z 
\frac{ {}_1F_1 \left(\frac{1}{2} -\frac{\nu}{2};
\frac{3}{2};\frac{z^2}{2} \right)}
{\Gamma \left(-\frac{\nu}{2} \right)} \right\}.
\label{hypergeometric}
\end{eqnarray}
This representation is valid along the entire real axis. Since it is simpler
than any form we have found in the literature, we briefly sketch its 
derivation. We start by observing that the Weber Eq.~(\ref{eqweber})
may be transformed to the confluent hypergeometric equation by changing to
the independent variable $s=z^2/2$ and the dependent variable
$\xi = e^{s/2} \chi_1$. By means of this 
transformation, which is valid for $z>0$, we obtain
a confluent hypergeometric equation~\cite[p. 671]{morse} with parameters
$a=-\nu/2$ and $b=1/2$.  Thus we obtain the solution to Eq.~(\ref{eqweber})
\begin{equation}
D_\nu(z)= 2^{\nu/2} e^{i \nu\pi /2} e^{-z^2/4} 
        {\cal U}_2 \left( -\frac{\nu}{2}, \frac{1}{2}, \frac{z^2}{2} \right)
\end{equation}
Here ${\cal U}_2$ is a confluent hypergeometric function of the third
kind~\cite[p. 612]{morse}, which is well-behaved as $z \rightarrow \infty$, 
and the constant pre-factors have been inserted to be
consistent with the usual convention~\cite[p. 1641]{morse}.
We may now use a well known joining formula~\cite[p. 673]{morse}
relating ${\cal U}_2$
to the standard confluent hypergeometric function ${}_1F_1$ to obtain
the final result~(\ref{hypergeometric}).

So far, our derivation applies only to $z>0$, but we can see from the
differential Eq.~(\ref{eqweber}) that $D_\nu (z)$ is analytic at $z=0$,
and, since the function ${}_1F_1$ is known to be analytic over the entire
complex plane, we conclude that Eq.~(\ref{hypergeometric}) holds along
the entire real axis. We note that the standard 
treatises~\cite{whittaker,morse} both arrive at the representation
Eq.~(\ref{hypergeometric}), valid for $z>0$, but offer a much
more complicated continuation of it along the negative $z$-axis
(see, for example, the asymptotic analysis in~\cite{whittaker}, sections
16.51 and 16.52, or the identical discussion by~\cite{morse} on
p. 1641).

\end{document}